\documentclass[journal]{IEEEtran}

\usepackage{bm,amsmath,amssymb,amsfonts,graphicx,multirow,color,colortbl,url}
\usepackage[linesnumbered,ruled,vlined]{algorithm2e}
\usepackage{cite}
\usepackage{subcaption,sidecap}
\usepackage[font={small}]{caption}
\usepackage{rotating}
\usepackage{mathtools}

\usepackage[table]{xcolor}

\newcommand{\argmin}{\arg\!\min}

\newcommand{\svm}{\text{\tiny{(svm)}}}
\newcommand{\mlp}{\text{\tiny{(mlp)}}}
\newcommand{\rnn}{\text{\tiny{(rnn)}}}
\newcommand{\groundtruth}{\text{\tiny{(gth)}}}

\newcommand{\FigsDir}{./Figs}

\makeatletter
\newcommand{\removelatexerror}{\let\@latex@error\@gobble}
\makeatother

\begin{document}
\title{An Intelligent Bed Sensor System for Non-Contact Respiratory Rate Monitoring}

\author{Qingju~Liu,
        Mark~Kenny,
       Ramin~Nilforooshan,
       Payam~Barnaghi,~\IEEEmembership{Senior Member,~IEEE,}
\thanks{Q. Liu is with the Centre for Vision, Speech and Signal Processing, University of Surrey, Guildford, GU2 9NS, UK.
e-mail: q.liu@surrey.ac.uk}
\thanks{M. Kenny and R. Nilforooshan are with Surrey and Borders Partnership NHS Foundation Trust, ACU, Holloway Hill, Chertsey, KT16 0AE, UK.
e-mail: \{mark.kenny, ramin.nilforooshan\}@sabp.nhs.uk}
\thanks{P. Barnaghi is with the Department of Brain Sciences at Imperial College London. e-mail: p.barnaghi@imperial.ac.uk}

\thanks{P. Barnaghi and R. Nilforooshan are also with Care Research and Technology Centre at the UK Dementia Research Institute.} 
}

\markboth{}%
{LIU \MakeLowercase{\textit{et al.}}: IoT-BASED NON-CONTACT RESPIRATION MONITORING}
%

\maketitle

\begin{abstract}
We present an IoT-based intelligent bed sensor system that collects and analyses respiration-associated signals for unobtrusive monitoring in the home, hospitals and care units. A contactless device is used, which contains four load sensors mounted under the bed and one data processing unit (data logger). Various machine learning methods are applied to the data streamed from the data logger to detect the Respiratory Rate (RR). We have implemented Support Vector Machine (SVM) and also Neural Network (NN)-based pattern recognition methods, which are combined with either peak detection or Hilbert transform for robust RR calculation. Experimental results show that our methods could effectively extract RR using the data collected by contactless bed sensors. The proposed methods are robust to outliers and noise, which are caused by body movements. The monitoring system provides a flexible and scalable way for continuous and remote monitoring of sleep, movement and weight using the embedded sensors.

\end{abstract} 


\IEEEpeerreviewmaketitle

\section{Introduction}
\label{sec:introduction}
Around 11\% of deaths in hospitals are due to failure to detect deterioration in patients conditions \cite{stats}. Respiratory rate (RR), combined with body temperature, blood pressure and pulse rate, are the four vital signs closely related to a person's health conditions, such as disease progression and clinical severity. In particular, it is reported in~\cite{Subbe} that RR is the best discriminator in identifying high-risk patient groups before adverse conditions such as cardio‐pulmonary arrest or critical care admission. Naturally, RR monitoring facilitates detection of respiratory-related symptoms such as sleep apnoea, chronic obstructive pulmonary disease and asthma~\cite{Dias}. Besides that, changes in RR also have a strong relationship with psychological aspects such as stress and anxiety~\cite{Suess,Masaoka} and cognitive load~\cite{Grassmann}. RR can also be linked to polysomnography since its changes reflect different stages of sleep \cite{Douglas,Redmond}. The current practice to monitor the respiratory rate in hospitals and care setting is usually based on manual observations by clinical staff. This hinders collecting continuous and reliable data at scale to develop early risk analysis and interventions. Developing automated and continuous RR monitoring solutions assists creating more robust, and scalable early interventions in adverse health conditions where RR can be a key indicator \cite{Cretikos}. 

Various RR estimation methods have been developed, as reviewed in \cite{Folke, Al-khalidi}. Overall the RR estimation methods can be categorised into two groups: contact and non-contact. In some contact methods sensors are attached near the airway or body parts (neck, chest, abdomen .etc) where the collected signals directly reflect the respiratory activities, e.g. airflow, breathing acoustic, lung movements. Other contact methods use sensors that carry implicit respiratory information, e.g. PPG and ECG devices that contain some form of respiratory modulation to the main component. A traditional and still commonly-used non-contact method requires trained clinical observations by counting chest rises during a specified period (e.g. a full minute)~\cite{Flenady}. The non-contact methods utilise sensors such as pressure \cite{Zhu}, optical imaging \cite{Nakajima}, thermal images \cite{Murthy,Lewis} and radar signals \cite{Rahman,Vinci} that could detect mechanical movements or heat maps caused by respiration. It is also pointed in \cite{Al-khalidi} that non-contact sensing methods provide improved patient comfort and accuracy since the ``distress caused by a contact device may alter the respiration rate''. 

Despite a large number of existing sensing techniques for RR monitoring, most of the systems mentioned above employ a single device, each requiring a trained technician/clinician to operate. The manual operation makes it challenging for large-scale real-time monitoring. Internet of Things (IoT) technologies can provide a solution to this limitation. An IoT system could have a complex structure such as sensor nodes, data storage, network, users. However, its concept is quite straightforward, i.e. a ``communicating-actuating network'' with a proliferation of sensing devices \cite{Gubbi}. 

We describe a non-contact IoT-based system for real-time in-bed RR monitoring. The proposed system consists of five main components: bed sensor devices, gateway, back-end, machine learning (ML) algorithms and user interface. Each device set contains four load sensors mounted beneath bed legs, wire-connected to a data-logger that calibrates, processes and transmits the measured data. The data package sent from the data-logger includes respiration-associated waveform, and other metadata such as the device id, patient id, weight, centroid, and posture. The gateway receives data over a wireless network and creates a common data format before storing it in a repository. With various configurations and specifications, the back-end coordinates database management and other web services such as message bus, authentication middleware and network protocols. The ML component estimates the real-time RR and the user interface visualises the data associated with the registered patients. The proposed system offers the following advantages over traditional RR extraction methods.  

\begin{itemize}
\item Low awareness of operation. 
It is reported in \cite{Hill} that the ``awareness of monitoring appears to reduce respiratory rate''. Our sensors mitigate this problem since they are non-contact and mounted under the bed, aiming to monitor respiration at rest or sleep. As a result, the person feels no disturbance due to the continuous operation. It also does not require postural or positional constraints, bringing great comfort to users.    

\item Upgradeable RR estimation methods. The gateway, the backend and the ML component run on a remote server, instead of a physical server at the monitoring site. The ML module runs continuously with but does not interfere with the data collection process. The deployed ML models can be re-trained and fine-tuned to adapt to new patterns and statistics in the collected data. 

\item Scalable and extensible ecosystem. New devices can be added and registered by authorised users via the user-interface, and be associated with assigned patients. The ML function can be expanded, with more customised features, e.g. to predict the sleep quality. Other types of devices, e.g. wearable, can be integrated into our system to collect physiological signals for comparison or fusion with the respiratory signals. 

\item Remote monitoring. Authorised users, e.g. clinicians, have access to the user interface and can remotely view and manage both real-time and historical records.    
 
\end{itemize}

The technical novelties of this work include \textit{i)} developing an end-to-end system which starts from raw sensory data collection to end-user interface and visualisation. \textit{ii)} providing a flexible IoT architecture that can include other sensors devices as well that facilitates great system flexibility. \textit{iii)} constructing a novel method to analyse the noisy and dynamic data to detect the respiratory rate from the unobtrusive sensory data and evaluating the work and comparing the performance with conventional benchmark algorithms.

The remainder of the paper is organised as follows. Details of the system are introduced in Section~\ref{sec:system}. The RR estimation methods are explained and justified in Section~\ref{sec:ML}, where various algorithms such as Support Vector Machines (SVM) and Deep Neural Network (DNN) of different architectures are combined with either Hilbert transform or peak counting for robust RR extraction.  Section~\ref{sec:exp} presents and analyses experimental results on real-world recordings. Discussions and insights for future work are given in Section~\ref{sec:conclusion}.

\section{The Proposed System}
\label{sec:system}
The proposed system extracts accurate real-time RR from the bed sensor device. The system components and the bed sensor unit are described below.

\subsection{Bed Sensor Unit}
\label{sec:bedsensor}

\begin{figure}[tbh]%
  \centering
    \begin{subfigure}[b]{0.45\textwidth}
        \includegraphics[width=\textwidth]{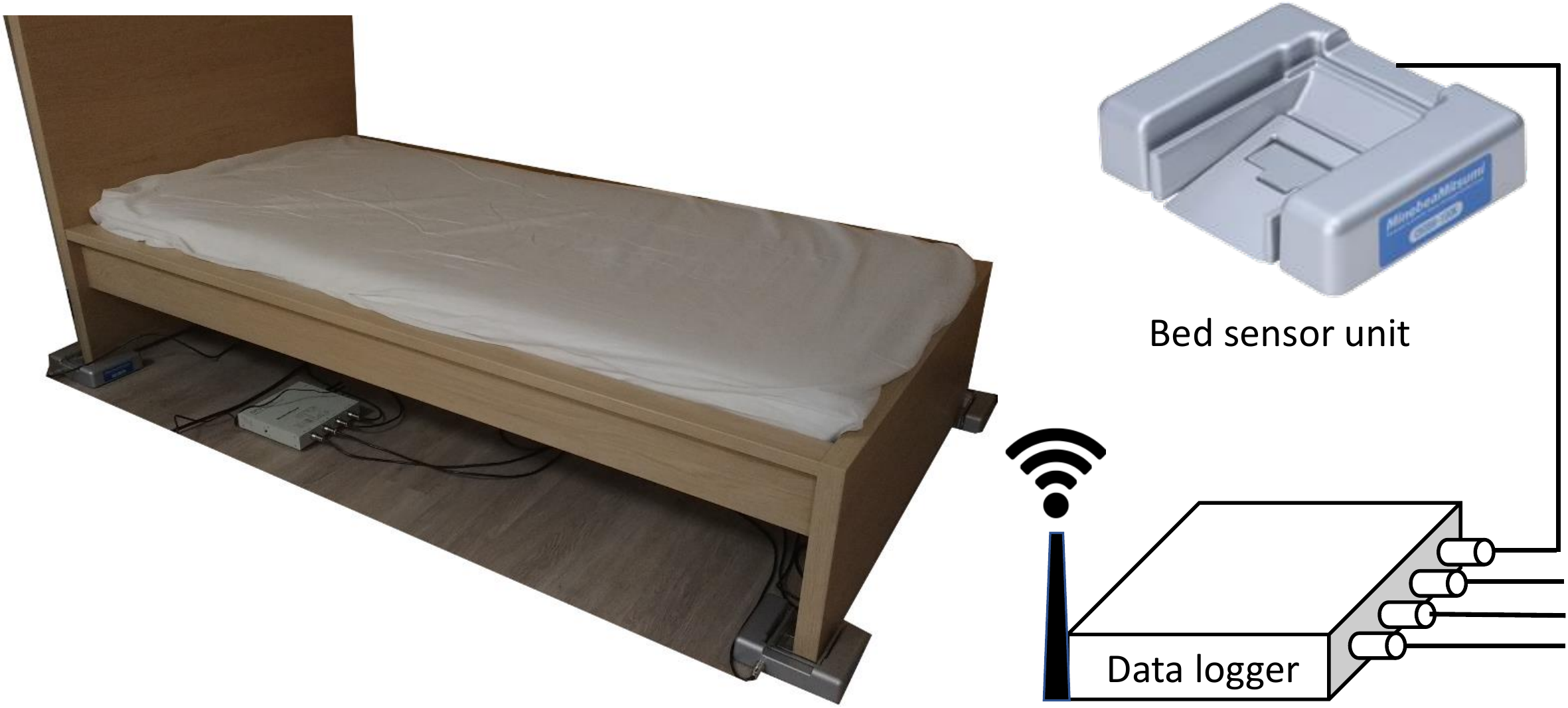}
        \caption{Bed sensor installation}
    \end{subfigure}
    
    \begin{subfigure}[b]{0.45\textwidth}
        \includegraphics[width=\textwidth]{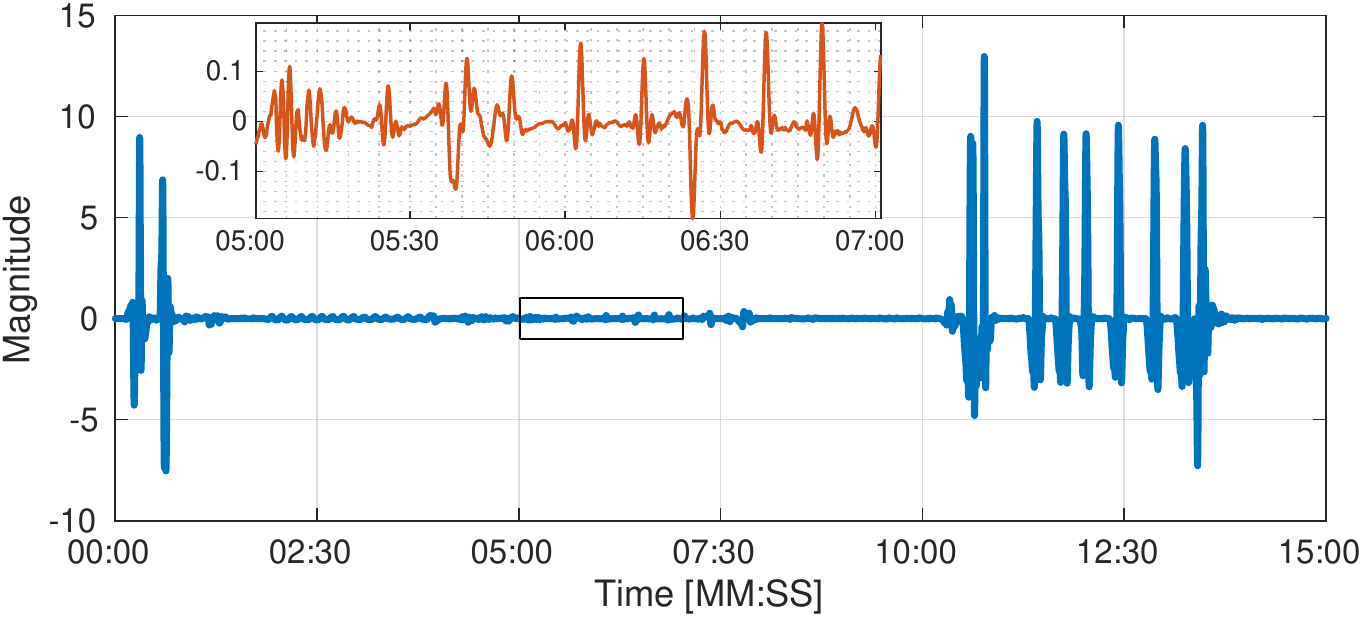} \\ \vspace{-0.22cm}
        \caption{Bed sensor waveform signal}
    \end{subfigure}
  \caption[]{A bed sensor unit and its wired connection to the data logger are shown (top). The data logger calibrates, pre-processes, and sends away data measurements. A signal snippet lasting of 15 minutes output by the bed sensor is plotted (bottom). The embedded plot zooms in a short segment highlighted in the small rectangle when the participant is not moving significantly, from which a periodic respiratory pattern is observed. }
  \label{fig:sensor}
\end{figure}

The bed sensor hardware and data-logger are developed by MinebeaMitsumi\footnote{https://www.eminebea.com/en/bedsensorsystem/}. A bed sensor device contains a data logger and four sensory units, which are essentially strain gauges (i.e. load cells). The sensor units have a high sensitivity. Thus any pressure change due to mechanical movements can be captured and reflected in the vibrations of the sensor measurement, including the very trivial breathing activity. The data logger calibrates the raw data and creates a waveform signal mirroring body movements. Other information such as body weight, centroid and posture can also be extracted from the data-logger. This additional information is not in the scope of this paper and will not be discussed further. A wireless data transmitter \& receiver is integrated into the data logger for data communications. A bed-mounted device and its output waveform signal are illustrated in Fig.~\ref{fig:sensor}.  

We need to stress that, the bed sensor system picks up vibrations caused by mechanical movements. Regular movements such as rollover, situp, coughing, will yield relatively large magnitude in the measured waveform, masking away the breathing information, as shown from the 10th minute in Fig.~\ref{fig:sensor} (bottom). However, when the participant is at rest, respiratory related movements can be captured by the measured waveform, despite exhibiting very small magnitude (two orders of difference as compared to that caused by regular movements). Yet, these ``weak'' signals contain rich respiratory information, from which the underlying RR can be estimated with appropriate signal processing techniques.

\subsection{The System Architecture}

\begin{figure*}[tb]%
  \centering
  \includegraphics[width=18cm]{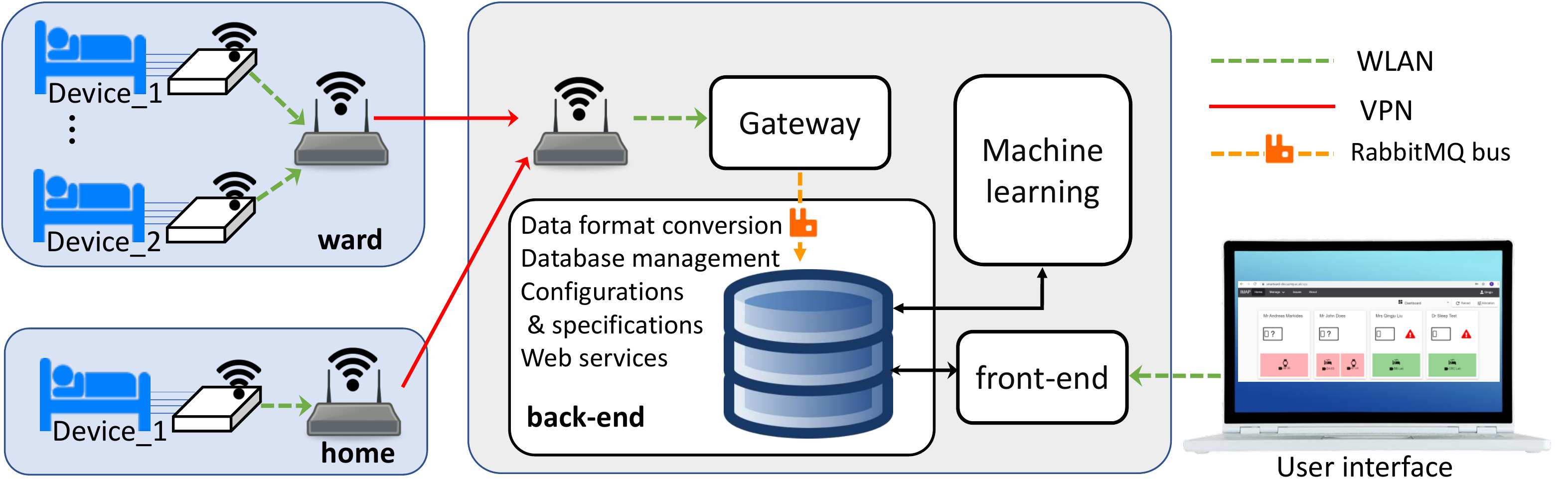}
  \caption[]{Schematic of the proposed RR monitoring system. Different components, i.e. local devices that collect patients' biological information, service providers (server), end-users, can either be distributed or share the same access point. The diagram illustrates a  more challenging distributed case in which the local devices and the server are connected by a virtual private network (VPN), while authorised access between the server and end-users are built, to protect data security and patient privacy. }
  \label{fig:IoTsystem}
\end{figure*}

The overall design of the system is shown in Fig.~\ref{fig:IoTsystem}, featuring a scalable architecture that enables real-time monitoring of multiple participants. \textit{Distributed} devices at different locations are connected to the central data processing point, which accommodates most of the compute-intensive and resource-demanding services, including the gateway, back/front-ends and the ML component. For simplicity, this central processing point will be referred to as the server hereafter. 

A set of back-end services run on the server to provide system efficiency, security, stability and thus high-quality user experiences. For instance, authentication services for user account management, device $\&$ patient services for registration and association, issue, comment and notification services to report issues and push front-end visualisation.  To support interoperability with other healthcare platforms, the data packages containing the collected data are transformed to Fast Healthcare Interoperability Resources (FHIR) \cite{fhir} messages.  To efficiently handle the large amount the data packages collected by the gateway from the distributed devices, a RabbitMQ \cite{rabbitmq} message broker queues and forwards the converted FHIR messages to the storage system. MongoDB \cite{mongodb} is used to manage the database, which contains several security measures for data protection. 

The front-end component allows (authorised) users to access an interactive web interface to: \textit{i)} manage user accounts, \textit{ii)} provide real-time monitoring and also access historical data, \textit{iii)} manage the device and patient records, e.g. addition and registration of a new patient, \textit{iv)} flag up warning signs when the system detects irregular patterns in a participant's measurements that might evoke activity/health risks.

Processing and storage capacity of the system can be boosted by cloud computing and big data analysis techniques on a \textit{cloud server}, which could significantly improve the system scalability, promoting its deployment feasibility in the wards and hospitals with a large population of patients. This also reduces the hardware cost of setting up and maintaining local physical servers for individual usage at home environments, e.g. for domiciliary care of elderly living alone. In addition, a cloud server also facilitates system maintainability (e.g. service upgrade) and maximises the stability (e.g. avoiding compatibility issues of a local server). 

 Once the data collection and storage components are provided, the data is analysed using analytical and ML methods. Various ML methods have been applied to different use-cases in this domain \cite{Mahdavinejad}. However, the existing techniques are not scalable for the dynamic and multivariate data in our system. We first discuss our proposed methods for RR estimation using the dynamic sensory data in our system and then in the evaluation section, compare the performance of our proposed system with the conventional benchmark algorithms.

\section{Machine Learning for RR Estimation}
\label{sec:ML}

As introduced in Section~\ref{sec:bedsensor}, the bed sensor generates a waveform that is respiration-related when the patient is at rest or approximately still. However, it is not practical to keep a patient still for a long time. Thus we need to mitigate the influence of outliers and unreliable data caused by body movements, while continuously monitoring the RR. 

\subsection{Data Reliability}

\begin{figure}[tb]%
  \centering
  \includegraphics[width=0.45\textwidth]{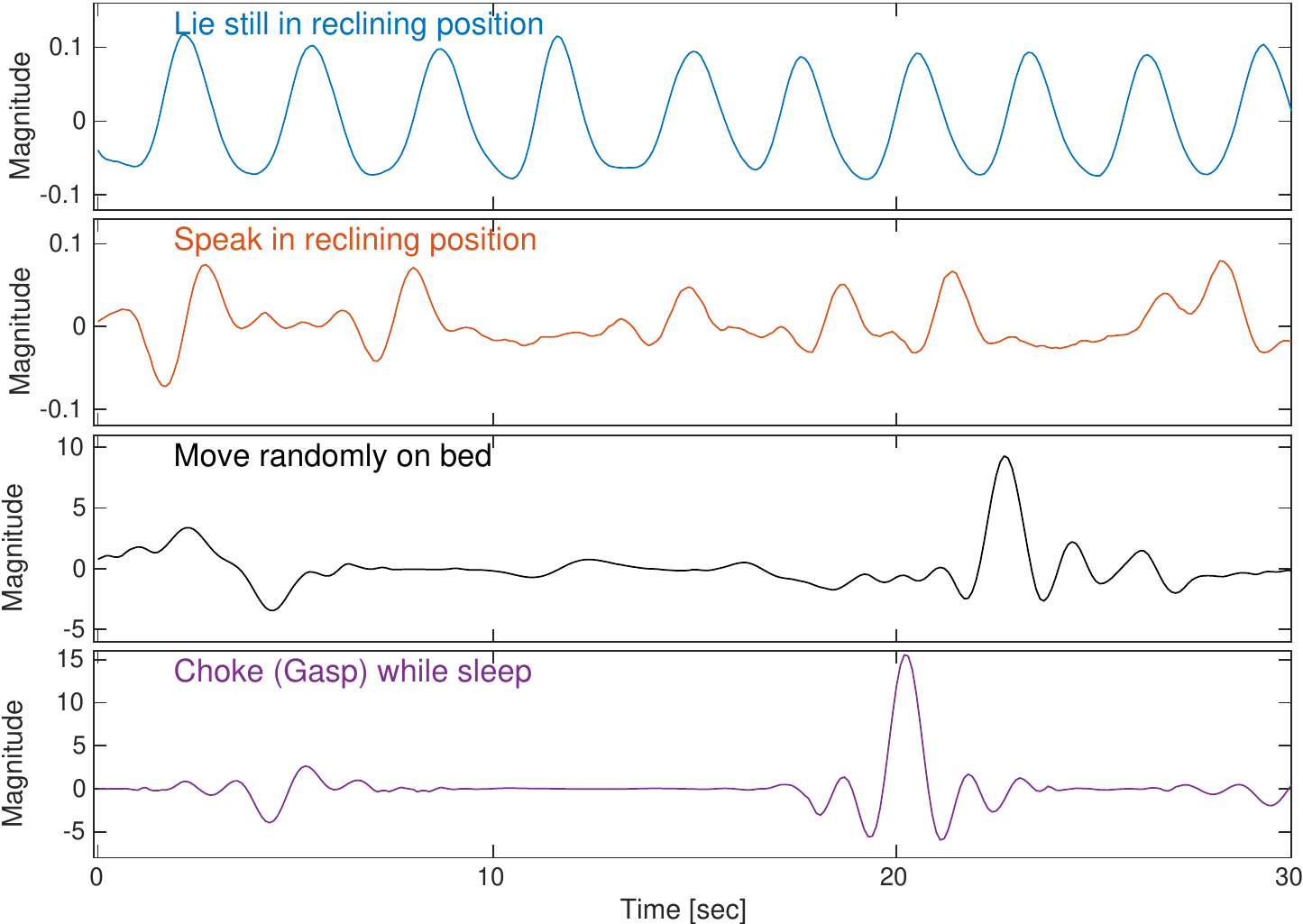}
  \caption[]{One-minute-long sensor waveform illustration, when the participant is engaged in four status, with (bottom three) and without (top one) body movements other than respiration. Body movements will modify and hide breathing activities in the waveform, resulting in unreliable data for RR estimation.}
  \label{fig:realsensorsignal}
\end{figure}

Fig.~\ref{fig:realsensorsignal} illustrates the collected bed sensor signals when the participant is engaged in four status: staying still in a reclining position, speaking in a reclining position, moving randomly, and gasping (the figures are shown from top to bottom). When the participant lies still, the signal yield sinusoidal patterns with a very small magnitude. Each peak-through, in this case, is associated with a full respiratory cycle. The RR can then be calculated directly, by counting peaks in a certain period.  

When there are other activities besides respiration such as speaking, the breathing-associated waveforms collected by the sensors can be significantly distorted. Speech breathing has irregular cycles due to changes of depth and duration of inhalation and exhalation \cite{Rochet-Capellan}. Speaking involves muscle activities of low-jaw, neck and shoulder, and thorax movements, and these kinematics are not negligible as compared to breathing activities, introducing artefacts and outliers. When relatively ``heavier''  movements are involved such as random movements and gasps, waveforms with large magnitudes are obtained, as shown in the last two subplots in Fig.~\ref{fig:realsensorsignal}. These movements exhibit much more complex and irregular patterns, due to the randomness in position, direction and force of the motion. Most importantly, respiratory information is distorted by these movements. So the affected sensory measurements might not be reliable for RR estimation. 

A solution to address the data reliability issue is to apply a classifier followed by the RR estimation. The classifier detects the reliability of a sample or a short segment of the sensory measurement, i.e. an accept/reject binary classifier. Afterwards, information contributed by the selected reliable data can be used to calculate RR robustly. 

Based on the empirical experience, body movements often result in larger magnitudes (higher energy) than breathing activities. One straightforward solution is to impose energy-thresholding to the waveform signal. This solution, however, cannot adapt well to distinguish patterns caused by breathing and very light movements such as speaking. In addition, it is difficult to define the threshold, considering individual differences in weight, position (edge or or centre point of the bed), posture (recline, lying, side-lying) and breath type (deep breath, light breath), which all affect the energy level of the signal. 

An alternative method is supervised training by providing sufficient labelled data and an appropriate classification model. A supervised algorithm can be applied to unveil the hidden structure and explore the underlying information in the data.  

\subsection{Supervised Training}

A variety of supervised classification methods can perform binary clustering, e.g. Logistic Regression (LR), decision trees, random forest, Support Vector Machines (SVM) \cite{Cortes}, and Deep Neural Network (DNN) models  \cite{LeCun}. Decision trees and random forest are often employed on data with multiple types of features. This, however, is not the case for our 1D time-series waveform. LR is prone to over-fitting and assumes a linear combination of the input features. On the other hand, SVM shows advantages of a rich hypothesis space with a large number of kernel functions that handle linearly-challenging data. Neural Network models are capable of ``discovering intricate structure in large datasets'' given a sufficient set of training data \cite{LeCun}. As a result, SVM and DNN are considered to build a robust classifier in this paper. 

We denote $s(t)$ as the waveform signal indexed by the time instance $t$, and $F_s$ as the sampling rate. We aim to build a classifier that detects the reliability of the $n$-th frame ${\bf s}(n) = [s((n-1)F_s+1),...,s(nF_s)]^T$, where the superscript $T$ represents the transpose. This ground-truth reliability label is denoted as $y(n)\in \{0,1\}$. To construct the inputs to the proposed classification models, the following features are extracted.

\subsubsection{Feature extraction} As mentioned earlier, breathing activities and other body movements yield different energy levels. To suppress unreliable outliers with large magnitudes and emphasise the more stable but ``weak'' respiratory waveform, we first normalise the signal to the range of
(-0.5, 0.5), with a scaled and shifted sigmoid function:
\begin{equation}
\bar{s}(t) = 1/\left(1+\exp(-\sigma s(t))\right)-0.5,
\end{equation}
where $\sigma$ is the scaling factor. The normalised $n$-th frame is denoted as $\bar{\bf s}(n)$.

To exploit the temporal information beyond the current frame $n$, e.g. the correlation between neighbouring frames, $2L+1$ frames centred at the $n$-th frame are extracted and concatenated together, resulting $(2L+1)F_s$-dimensional data: 
\begin{equation}
 {\bf z}(n) = [\bar{\bf s}(n-L)^T,...,\bar{\bf s}(n+L)^T]^T. 
\end{equation}
To avoid over-fitting of high-dimensional data, principal component analysis (PCA) is applied to reduce the data dimension. The PCA method generates the feature ${\bf x}(n)$:
\begin{equation}
 {\bf x}(n) = {\bf W}{\bf z}(n), 
 \label{eq:pca}
\end{equation}
where ${\bf W} \in \mathbb{R}^{Q \times (2L+1)F_s}$ contains $Q$ eigenvectors corresponding to the largest $Q$ eigenvalues, spanned by the eigenspace of all the $N$ training samples $\{{\bf z}(n)\}, n=1,2,...,N$. 

In principle, other aggregation techniques that efficiently represent the time-series can also be used for feature extraction. For instance, the discrete cosine transform and eigenvalues characterisation of the segmented data can also be used for dimensionality reduction \cite{GonzalezVidal}. However, in our data, we aim to implement a (near-) real-time method and avoid using the methods that have higher computational complexity. The extracted features preserve and represent the key characteristics of the data and facilitate constructing a more robust classifier.

\subsubsection{SVM} The extracted features are fed into the SVM training process, where the c-support vector classification is employed, which aims to optimise \cite{Chang}:

\begin{equation}
  \begin{split}
   &\argmin_{{\bf p}, p_0, \xi} = 0.5 \| {\bf p} \|^2 + C\sum_{n=1}^N \xi_n  \\
  \text{ s.t. \hspace{0.5cm}} & y(n)({\bf p}^T \phi({\bf x}(n)) + p_0) \geq 1-\xi_n, \xi_n>=0,
   \end{split}
\label{eq:svm}
\end{equation}
where $\|\cdot\|$ calculates the Frobenius norm; $C$ is the box constraint that weights the penalty term of slack variables $\xi_n$; $\phi(\cdot)$ applies conversion of the input data to a higher-dimensional space. Eq.~(\ref{eq:svm}) is solved by a dual optimisation problem defined on the kernel space, where the kernel is the inner product of two converted variables. The kernel can be selected from a set of kernel functions. The Radial Basis Function (RBF) kernel was chosen based on the pilot classification tests. The dual optimisation process is regularised by the box constraint $C$ and a kernel scale parameter $\gamma$, which can be set via a heuristic procedure such as grid searching once the kernel function is chosen. The trained model, $\Psi^{\svm}(\cdot)$ is applied on a data point for reliability detection:

\begin{equation}
 {\hat y(n)} = \text{sign}\left(\Psi^{\svm}({\bf x}(n))\right).
\end{equation}
When a bagged ensembles of SVM models $\Psi_k^{\svm}(\cdot), k = 1,...,K$ are trained via $K$-fold cross validation, the final prediction becomes:
\begin{equation}
 {\hat y(n)} = \text{sign}\left(\sum_{k=1}^K \Psi_k^{\svm}({\bf x}(n))\right).
\end{equation}

\subsubsection{DNN}
In recent years, DNNs have gained increasing popularity in many research fields, showing dramatic improvements over conventional approaches in areas such as natural language processing \cite{Collobert}, speech recognition \cite{Hinton}, computer vision \cite{Krizhevsky} and control theory \cite{Volodymyr}. DNNs can learn the complex underlying structures in the data. We first implemented a conventional Multi-Layer Perceptron (MLP) model \cite{Malsburg} to construct the data reliability classifier. 

\begin{figure}[tb]%
  \centering
  \includegraphics[width=0.45\textwidth]{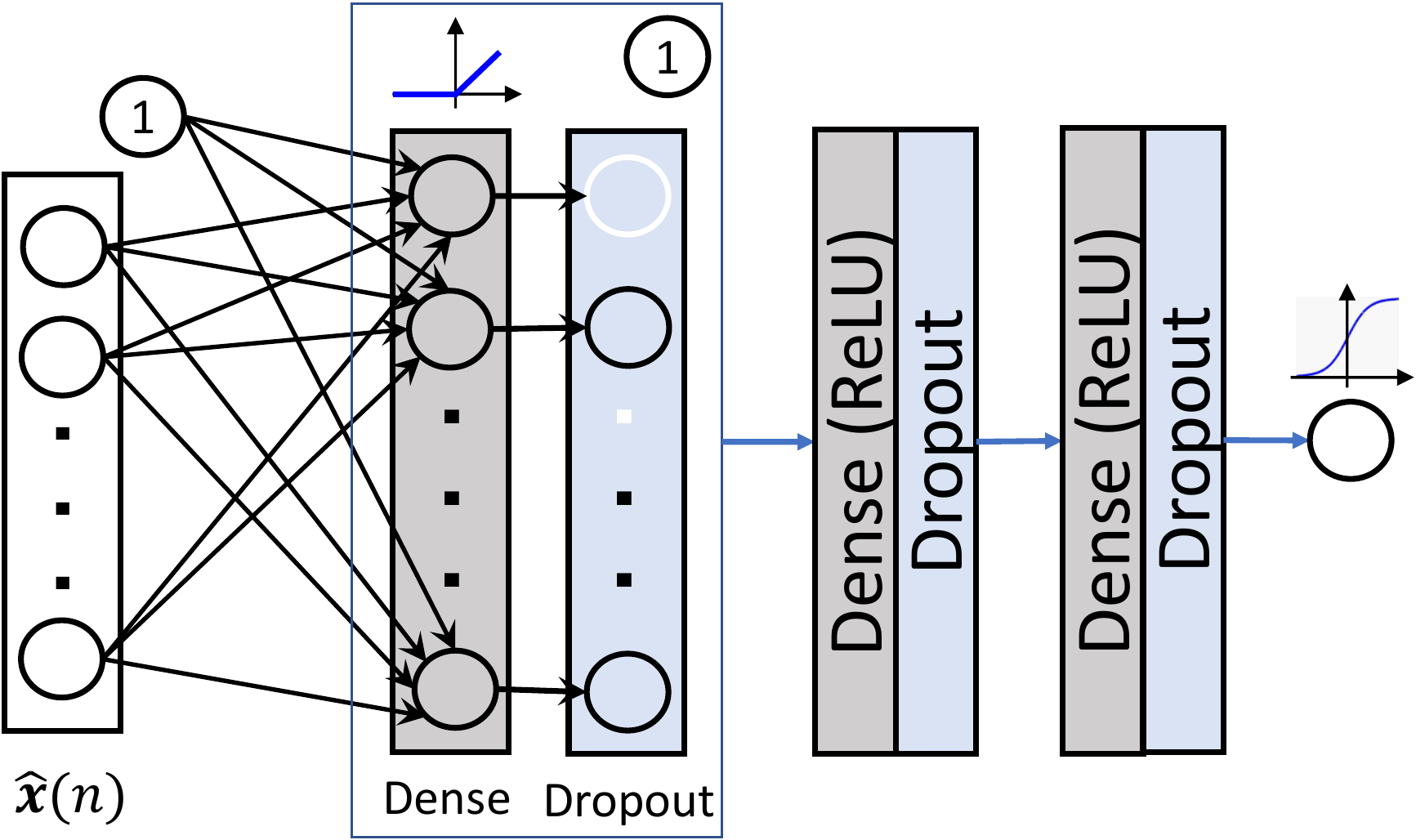}
  \caption[]{The implemented MLP network contains three hidden layers, each followed by dropout. Black lines between the input layer and the first hidden layer denote the dense combination, where the bias term is added via a unit of 1. The following layers and their dense connections are simplified for illustration purposes. Non-linear neurons are employed after the fully-connected dense layer, e.g. ReLU and sigmoid functions plotted on the top. 
  }
  \label{fig:mlp}
\end{figure}

MLP, also referred to as Vanilla Network, has a simple feedforward structure, as illustrated in Fig.~\ref{fig:mlp}. Units in two neighbouring layers are linked by fully-connected dense connections, followed by a nonlinear activation function. In our implementation, rectified linear units (ReLU) are used on hidden layers for fast gradient calculation in the backpropagation and quick convergence \cite{Glorot}. The sigmoid neuron is employed on the final output layer to constrain the prediction within the range of (0,1). To avoid over-fitting, dropout \cite{Srivastava} is employed after each dense layer, which randomly drops units together with their connections at a defined ratio. 

The same input feature for the SVM classifier, i.e. the PCA component from $2L+1$ concatenated frames in Eq.~(\ref{eq:pca}), is fed into the MLP model $\Psi^{\mlp}(\cdot)$ to obtain the final prediction:
\begin{equation}
 {\hat y(n)} = \text{sign} \left(\Psi^{\mlp}({\bf x}(n)) - 0.5 \right).
\end{equation}

To train the MLP model $\Psi^{\mlp}(\cdot)$, the cross-entropy is minimised as the loss function during the backpropagation:
\begin{equation}
-y\text{log}(\Psi^{\mlp}({\bf x})) - (1-y)\text{log}(1-\Psi^{\mlp}({\bf x})),
\end{equation}
where the time index $n$ is omitted for simplicity. 

\begin{figure}[tb]%
  \centering
  \includegraphics[width=0.48\textwidth]{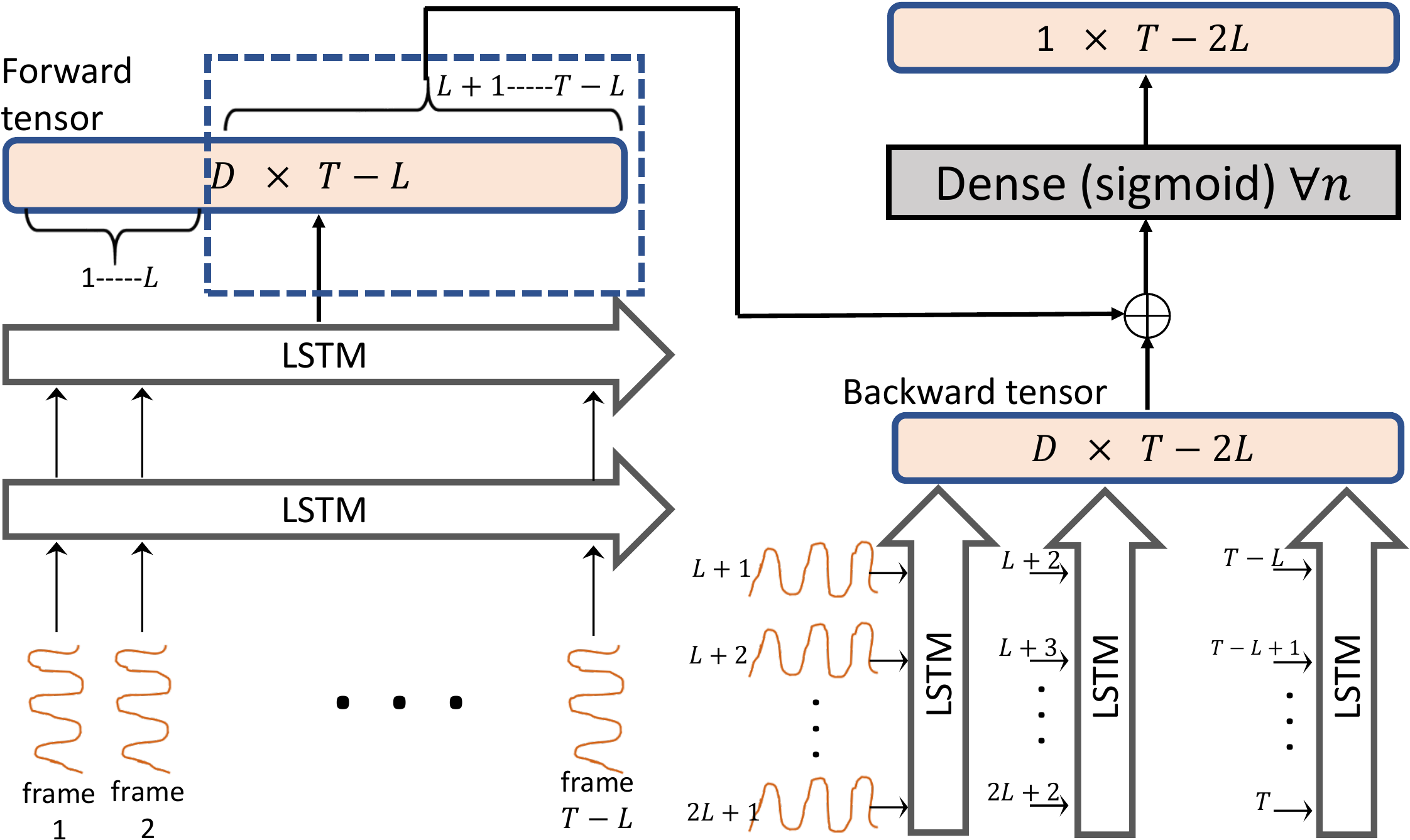}
  \caption[]{The implemented hybrid RNN network contains two stacked layers of LSTM, and one backward LSTM that exploits future $L$ frames, outputting hidden forward and backward tensors, which are fused by summation. A fully-connected dense layer with sigmoid activity function is applied at each time instance of the fused tensor, to generate the consecutive predictions. The LSTM layers are illustrated with the thick hollow arrows, and the pink blocks denote the tensor outputs.   
  }
  \label{fig:rnn}
\end{figure}

The straightforward MLP structure, however, has the limitation that its receptive field is constrained by the fixed input size, e.g. $2L+1$ frames. Thus the long-temporal information is not exploited. To compensate for the negligence of long-term context information, we have implemented another hybrid network mainly based on the Recurrent Neural Networks (RNN) \cite{Elman}. The hybrid RNN structure is illustrated in Fig.~\ref{fig:rnn}. 

RNN can take input of varying length of $T$, and we have $T \geq 2L+1$ frames of input data. Instead of using the extracted features such as the PCA component employed in SVM and MLP methods, we use the normalised time series $\bar{\bf s}(n), n=1,...,T$ directly.  RNN layers with Long Short Term Memory (LSTM) \cite{Gers} units are utilised, which explore the historical information from all previous frames when predicting the current frame. An alternative is bidirectional \cite{Schuster} LSTM (BiLSTM) that explores both previous context and future context to learn the time-dependencies for the current frame. BiLSTM, however,  results in the latency problem and requires a whole sequence to run, which is not suitable for online processing. To address this issue, we have implemented a hybrid structure, as shown in Fig.~\ref{fig:rnn}. The proposed model modifies the mechanism of bidirectional recurrent networks, while also offers the opportunity for online processing at a tolerable latency. To provide a consistent base for comparison with SVM an MLP model which utilise both previous and future $L$ frames, the proposed RNN structure also has $L$ frames of latency.  

The normalised time series $[\bar{\bf s}(1),...,\bar{\bf s}(T-L)]$ first goes through two staked LSTM layers that both contain $D$ filters, resulting a forward tensor of size $D \times (T-L)$. In parallel, the time series $[\bar{\bf s}(L+1),...,\bar{\bf s}(T)]$ are segmented into $T-2L$ overlapped segments each containing $L+1$ frames. Each segment goes through a backward-directional LSTM layer that contains $D$ filters, and only the last frame output is reserved. Outputs from all segments are concatenated to form a backward tensor of size $D \times (T-2L)$. The forward and backward tensors are added together, and to keep the dimensional consistency, only the last $(T-2L)$ frames of the forward tensor are fused with the backward tensor. Sigmoid units are applied to the fused tensor at each frame, generating $T-2L$ label estimations associated with the $T-2L$ frames indexed from $L+1$ to $T-L$.

To apply this RNN network to an online prediction scenario (i.e. generating a label prediction when a new frame arrives), $L$-second-delay is allowed to obtain the future $L$ frames for the backward tensor update. In addition, hidden outputs of the two stacked LSTM layers at the previous frame are kept representing historical information. Denoting the RNN model as $\Psi^{\rnn}(\cdot)$, the reliability is obtained as:

\begin{equation}
 {\hat y(n)} = \text{sign} \left(\Psi^{\rnn}([{\bf {\bar s}}(1),...,{\bf {\bar s}}(n+L)]) - 0.5 \right).
\end{equation}

\subsection{RR Estimation}
After applying the converged classification models to the collected waveform to detect whether a frame of data is reliable or not, RR can then be extracted from these reliable frames. Two RR estimation methods are implemented, one based on peak counting, and the other Hilbert transform (HT). 

\subsubsection{Peak counting} 
From the top row in Fig.~\ref{fig:realsensorsignal}, it can be observed that a full respiratory circle has a peak and a through. As a result, the number of breaths in a certain period is equivalent to counting peaks. 

Suppose $I$ segments of consecutive data is detected as reliable during a period of time, e.g. one minute, and $t_{i,1}$ to $t_{i,2}$ are respectively the starting index and end index of the $i$-th segment. This segment is denoted as ${\bf s}_i = [s(t_{i,1}),...,s(t_{i,2})]^T$. Peaks can be detected as the local maxima in the curve. To avoid finding false positives, constraints such as magnitude threshold and height difference threshold with the neighbouring through should be enforced. Assuming in total $J_i$ peaks are detected in ${\bf s}_i$, located at $p_{i,j},j=1,...,J_i$. The RR can be calculated as:
\begin{equation}
 \text{RR} = \frac{\left( \sum\limits_{i=1,J_i>1}^{I}(J_i-1) + \sum\limits_{i=1,J_i\leq1}^{I}J_i \right) 60/Fs}
 {\sum\limits_{i=1,J_i>1}^{I}(p_{i,J_i}-p_{i,1}) + \sum\limits_{i=1,J_i\leq1}^{I}(t_{i,2}-t_{i,1}) }.
 \label{eq:peak}
\end{equation}

\subsubsection{Hilbert Transform (HT)}
\begin{figure}
  \includegraphics[width=0.24\textwidth]{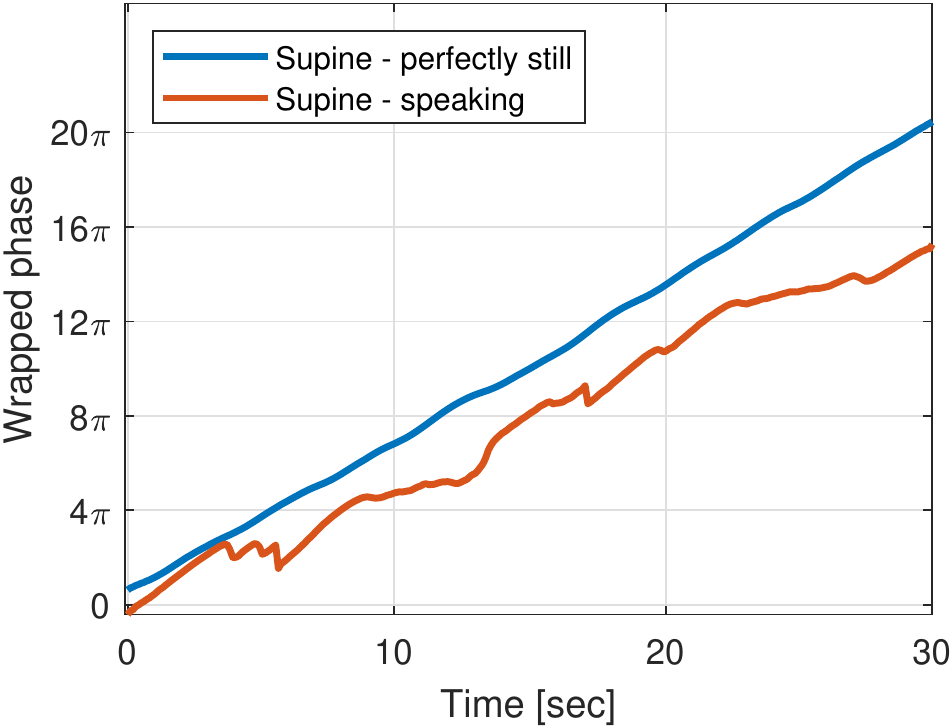}
  \includegraphics[width=0.24\textwidth]{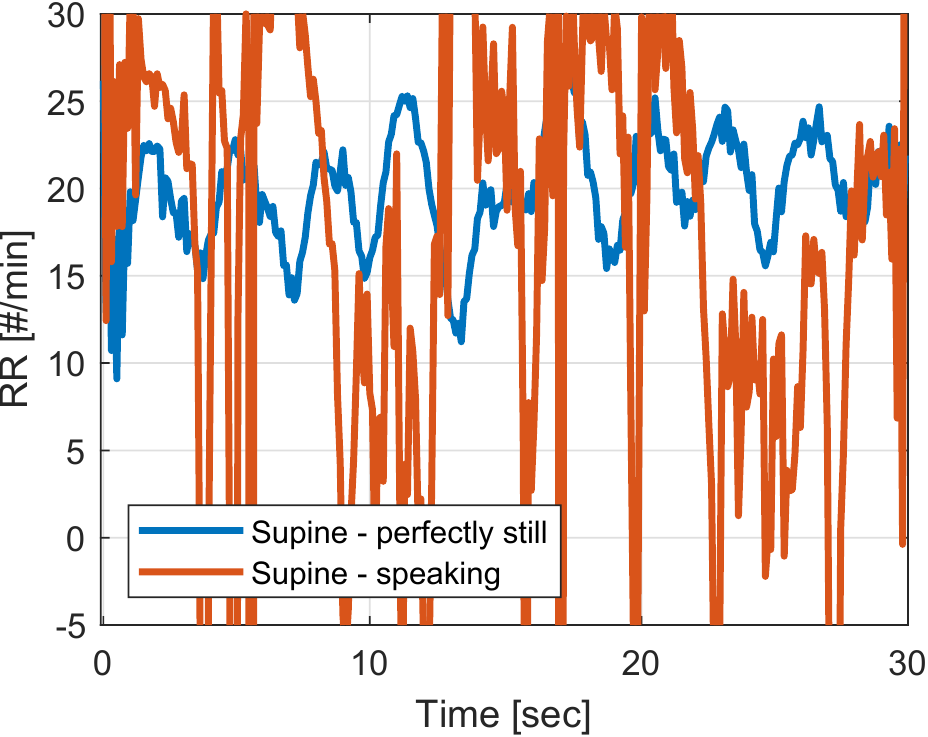}
  \caption{Applying HT to the signals in the top two rows of Fig.~\ref{fig:realsensorsignal} to obtain the wrapped phase (left), from which the IF can be calculated and mapped to RR (right). Using the HT-based slope fitting method, the RR values of 20 and 16 were calculated on these two sequences, respectively. } 
  \label{fig:ht}
\end{figure}

An HT method can be used for Instantaneous Frequency (IF) calculation, which is the gradient of the analytic signal's unwrapped phase. 
\begin{equation}
 \text{IF} = \nabla \text{U} \left( \angle {\cal H}(\cdot) \right), 
\end{equation}
Where $\cal{H}(\cdot)$ denotes HT, $\text{U}(\cdot)$ unwraps the phase, and $\nabla$ calculates the gradient. For our case, the IF can be directly mapped to RR by:
\begin{equation}
 \text{RR} = \frac{60}{2\pi}\text{IF}.
\label{eq:map}
\end{equation}
However, the IF is very sensitive to noise. A person's respiratory pattern is not an ideal sinusoidal wave. Thus local changes in inhalation and exhalation would cause large variations in the detected RR, as illustrated in Fig.~\ref{fig:ht}. The IF is essentially the slope of the phase curve at each time instance. However, if we look at the overall slope, it is quite smooth and reflects more stable respiratory information. As a result, a robust RR value can be estimated from all reliable data points by line fitting. 

Denote ${\bm \alpha}_i$ as the unwrapped phase from ${\cal H}({\bf s}_i)$:
\begin{equation}
 {\bm \alpha}_i = \text{U} \left( \angle {\cal H}({\bf s}_i) \right)  =  [\alpha(t_{i,1}),...,\alpha(t_{i,2})]^T,
\end{equation}
the slope fitting using contributions from all $I$ segments can be formulated as:
\begin{equation}
\underbrace{
  \begin{bmatrix}
    {\bm \alpha}_1 \left\{
    \begin{array}{c}
    \alpha(t_{1,1})\\
    \alpha(t_{1,1}+1)\\
    \vdots\\
    \alpha(t_{1,2})
    \end{array}
    \right.
    \\
    \vdots
    \\
    \vdots
    \\
    {\bm \alpha}_{I} \left\{
    \begin{array}{c}
    \alpha(t_{I,1})\\
    \alpha(t_{I,1}+1)\\
    \vdots\\
    \alpha(t_{I,2})
    \end{array}
    \right.
 \end{bmatrix}
 }_{{\bm \eta}}
 \approx
\underbrace{
 \begin{bmatrix}
    1       	&   1       &   0       & \dots     &   0       \\
    2       	&   1       &   0       & \dots     &   0       \\
    \vdots  	&  \vdots   &   \vdots  &   \vdots  &   \vdots  \\   
    \Delta t_1  &   1       &   0       & \dots     &   0       \\
    1       	&   0       &   1       & 0         &   \dots   \\
    2       	&   0       &   1       & 0         &   \dots   \\
    \vdots  	&  \vdots   &   \vdots  &   \vdots  &   \vdots  \\   
    1       	&   0       &   \dots       & 0     &   1       \\
    2       	&   0       &   \dots       & 0     &   1       \\
    \vdots  	&  \vdots   &   \vdots  &   \vdots  &   \vdots  \\   
    \Delta t_I  &   0       &   \dots       & 0     &   1       \\
 \end{bmatrix}
  }_{{\bm \Phi}}
\underbrace{
 \begin{bmatrix}
    b_0       \\
    b_1     \\
    \vdots  \\
    b_I     \\
  \end{bmatrix},
  }_{\bf b}
\label{eq:ht}
\end{equation}
where $\Delta t_I = t_{i,2}-t_{i,1}+1$ and ${\bf b}$ contains the overall slope $b_0$ and a bias term $b_i$ for each of the $i$-th segment, whose closed-form solution is:
\begin{equation}
 {\bf b} = ({\bm \Phi}^T{\bm \Phi})^{-1}{\bm \Phi}^T{\bm \eta}. 
\end{equation}
Integrating the sampling frequency that is omitted in Eq.~(\ref{eq:ht}), $\text{RR} = \frac{60Fs}{2\pi}b_0$ is obtained using Eq.~(\ref{eq:map}).

\section{Experiments}
\label{sec:exp}

\subsection{Data Recording}

To train and validate our proposed ML methods for RR estimation, data were recorded from four participants, each lasting around one hour. The data were collected on a standard single bed, as shown in Fig.~\ref{fig:sensor}.  A trained observer counted the number of breaths during specific periods when the breathing activity was visible. The observation information was used as ground-truth RR to validate our proposed method. In total, 56 minutes were successfully logged with a ground-truth RR. Details of the four participants, including gender, age, weight and Body Mass Index (BMI) are summarised in Table~\ref{tab:person}. None had any known respiratory conditions that would interfere with the ventilation. 

\begin{table}[!ht]
\caption{Summary of the participants' demographic information} 
\centering
\resizebox{0.9\columnwidth}{!}
{
  \begin{tabular}{lcccc} \hline\hline
  \rowcolor{lightgray} $\#$ & Gender & Age & Weight [kg] & BMI \\
  \hline
  \multicolumn{4}{l}{~} \\
  1     & M	&25	& 93 		& 24.26 	\\ 
  2 	& F 	&23	& 50 		& 19.53  	\\
  3	& M 	&46	& 68 		& 21.4 	 \\
  4	& F 	&25	& 62 		& 21.9        \\ 
  \hline\hline
  \end{tabular}
}
\label{tab:person}
\end{table}

To simulate breathing activities, as well as other activities and factors that might disturb the respiratory pattern, all participants were required to do the following 17 actions each lasting two minutes: lie in reclining position, lie in prone position, lie on the left side of body, lie on the right side of body, reclining position near the right bed edge, reclining position near the left bed edge, move randomly on bed, flounce legs, coughs, deep breaths, choke (gasp), speak, sit up on bed centre, sit on bed edge, rapid shallow breaths, body tremor, leg movements. 

In total, 14078 seconds of data were recorded. To train the reliability classifiers aforementioned in Sec.~\ref{sec:ML}, we manually labelled the data reliability per second (frame). Three classifiers were trained, as detailed below. 

\subsection{Building Data Reliability Models}
\label{sec:exp:models}
\subsubsection{SVM}
A SVM classifier was trained first. To extract features used in the SVM model, the scaling factor $\sigma$ in Eq.~(\ref{eq:map}) was empirically set to 10. Different temporal parameter $L$ were tested. PCA was applied to preserve the largest eigenvalues that predominate $95\%$ eigenspace. $14078-8L$ feature vectors were obtained in total, which were split to train-test ($80\%-20\%$). 10-fold ($K=10$) validation was employed to generate an ensemble of SVM models $\Psi_k^{\svm}(\cdot), k = 1,...,K$. 

To choose an appropriate kernel function, we ran pilot tests comparing linear functions, Gaussian radial basis functions and polynomial functions. The results showed better performance by using a Gaussian kernel (i.e. RBF kernel) in terms of prediction accuracy and therefore chosen hereafter. This is also in line with our intuition about the RR data, which seems to have a multivariate Gaussian distribution. Parameters $\gamma$ and $C$ were selected via grid searching, which varied with the temporal parameter $L$. 

We have tested five candidate $L$, and evaluated the performance on the test dataset, with the following three metrics: the overall detection error, precision=$\frac{\#\mbox{tp}}{\#\mbox{tp} + \#\mbox{fp}}$, and recall=$\frac{\#\mbox{tp}}{\#\mbox{tp} + \#\mbox{fn}}$. The results are summarised in Table~\ref{tab:eva}.

\begin{table}[th]
\centering
\caption{Quantitative evaluations of the built SVM models}
\resizebox{\columnwidth}{!}
{
  \begin{tabular}{cccccccc}
  \hline 
  {\multirow{2}{*} {$L$}} & \multicolumn{3}{c}{Parameters} & \multicolumn{1}{c}{ } & \multicolumn{3}{c}{Evaluations}  \\ 
  {}    & \multicolumn{1}{c}{$Q$} & {$C$} & {$\gamma$} & { } & {error} & {precision} & {recall} \\ 
  \cline{2-4}\cline{6-8}
  \multicolumn{8}{l}{~} \\
  0 & {3} & {2} & {0.05} & { } & {0.23} & {0.65} & {0.71} \\
  1 & {7} & {2} & {0.2} & { } & {0.14} & {0.80} & {0.78} \\
  2 & {11} & {2} & {0.3} & { } & {0.12} & {0.83} & {0.79} \\
  \rowcolor{lightgray}3 & {15} & {2} & {0.4} & { } & {0.11} & {0.87} & {0.78} \\
  4 & {19} & {2} & {0.5} & { } & {0.10} & {0.88} & {0.79} \\
  5 & {23} & {2} & {0.6} & { } & {0.11} & {0.88} & {0.77} \\
  \hline 
  \end{tabular}
}
\label{tab:eva}
\end{table}

As can be found in Table~\ref{tab:eva},  the performance improves with the increase of the temporal length and remains approximately stable after a specific length, e.g. 7 frames ($L=3$). However, a large $L$ also introduces a higher computational complexity due to the relatively high-dimensional data as well as a large number of support vectors in the model, making it prone to over-fitting. Thus we chose to use the model trained with $L=3$, as highlighted in the table. For fair comparisons, the same temporal length was also used for DNN-based models.    
 
\subsubsection{DNN}
We have implemented two DNN topologies: MLP and hybrid RNN structures. 

The same feature used for SVM training were also used in the MLP model. As illustrated in Fig.~\ref{fig:mlp}, the MLP model has 3 hidden layers. We set the size to 100 for each of the fully-connected layers, followed by a dropout of 0.2. The $l_2$ regularisation term was added with a weight of $10^{-6}$.  $20\%$ of the training data was used for validation. The learning rate was initialised to 0.001, with a decay rate of 0.99 after each epoch. The Adam optimiser \cite{Ba} was employed in the backpropagation. Mini-batches of size 64 were used, with each minibatch $\in \mathbb{R}^{64 \times Q}$, where $Q=15$ for $I=3$ situations. In total 1000 epochs were run, and the training data were randomised after each epoch. 

For RNN, the normalised raw data was used directly instead of the extracted features via the PCA representation in Eq.~(\ref{eq:pca}). As shown in Fig.~\ref{fig:rnn}, two layers of stacked LSTM layers were employed, with each layer of size 50, resulting in a forward tensor of dimension $D=50$ for each time instance. The backward LSTM also had a size of 50 for consistency. The beginning $L=3$ frames of the forward tensor were removed, before being added with the backward tensor. Mini-batches of size 32 were used, with each mini-batch $\in \mathbb{R}^{32 \times F_s \times T}$ where $20 \leq T \leq 30$. The other parameter setup is consistent with the MLP model.  

\begin{figure}[tb]%
  \centering
  \includegraphics[width=0.48\textwidth]{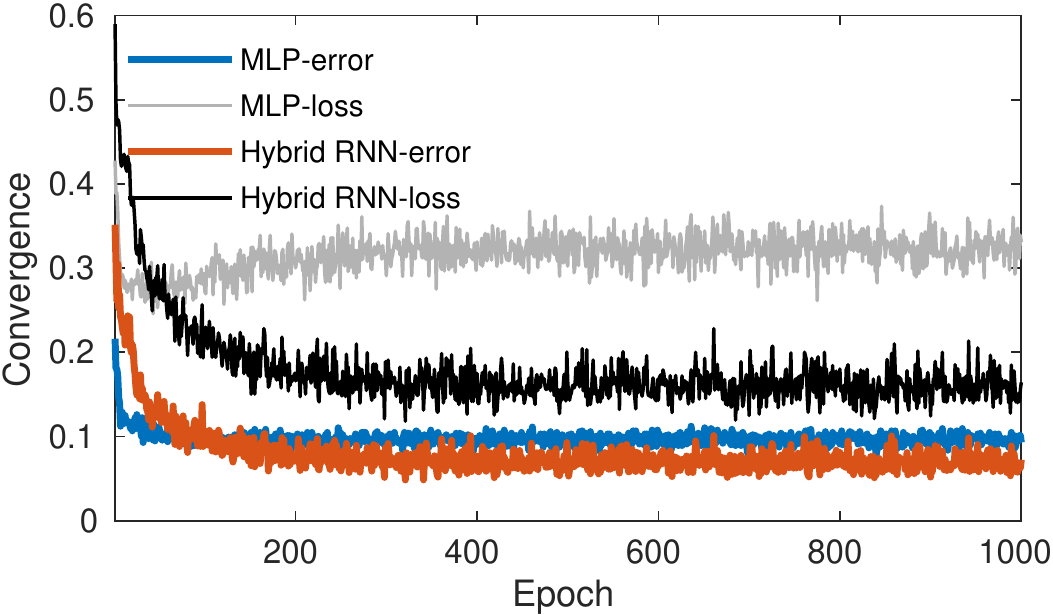}
  \caption[]{Convergence of the MLP and RNN methods on the validation dataset. The hybrid RNN model did not converge as fast as MLP. Yet, it gradually outperformed MLP after around 100 epochs with a stable and lower loss. Moreover, the hybrid RNN model also yielded a consistently higher accuracy, i.e. a lower error rate, than MLP.   
  }
  \label{fig:convergence}
\end{figure}

Fig.~\ref{fig:convergence} shows converge of the two DNN-based models on the validation data, in terms of the objective loss function (cross-entropy+regularisation terms), as well the overall prediction error. It can be observed that MLP converged quickly after around 20 epochs. Afterwards, its loss slightly degraded, and then remained smooth. RNN, on the other hand, showed relatively slow yet steady improvement in both loss and accuracy, exhibiting better performance than MLP, especially for loss convergence. The RNN model converged after around 400 epochs. 

Applied on the testing sequences, MLP showed an error of 0.10, precision of 0.88 and recall of 0.82, which were slightly better than SVM in terms of accuracy and precision, and relatively larger improvement of 0.04 in the recall. The RNN model showed advantages with a much larger margin, yielding an error of 0.03, precision of 0.98 and recall of 0.94. RNN had a much higher accuracy on testing sequences than the training dataset, for the following reason. 

In the RNN training stage, the forward temporal information was only used from short sequences lasting $20 \leq T \leq 30$ frames. The first few frames were more likely to suffer from a higher detection error due to the lack of long temporal dependencies. At the inference stage, the trained RNN model was applied to much longer sequences, where most frames (except the first few frames) benefited from accumulated temporal dependencies, thus yielding a higher accuracy. 

\begin{figure}[tb]%
  \centering
  \includegraphics[width=0.48\textwidth]{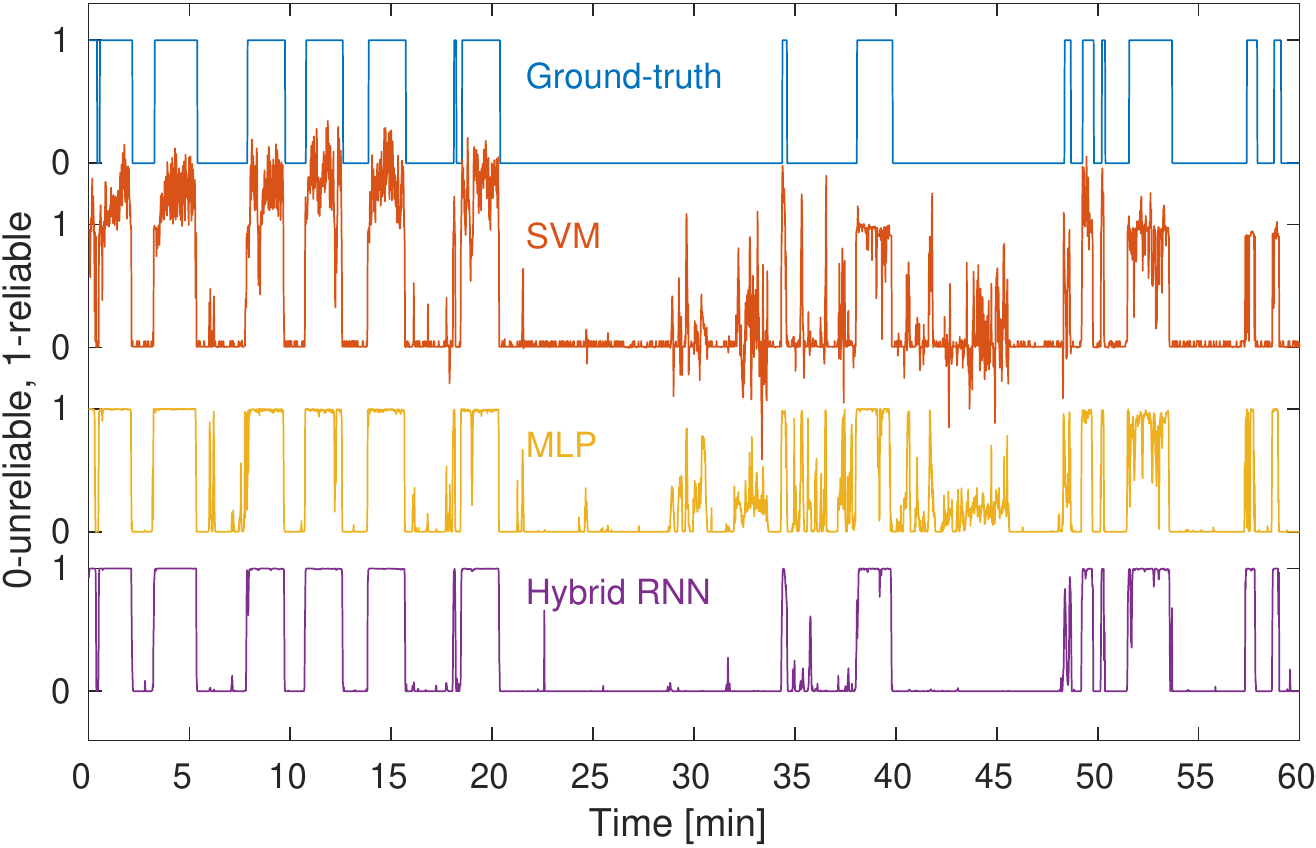}
  \caption[]{Comparison of prediction results (before the sign operator) from the three trained models with the ground-truth. The SVM prediction scores were linearly mapped to a range of [0 1], to be consistent with other methods.  
  }
  \label{fig:classifiers}
\end{figure}
To demonstrate the performance difference of the three reliability estimation models, we plotted an example of applying these models on a sequence lasting one hour, where the direct prediction results before the sign operator were shown. Overall, SVM and MLP showed a similar performance, and both suffered from uncertainties, especially from around the 30th to the 45th minutes. However, the hybrid RNN had a higher performance which is also close to the ground-truth. This observation was consistent with the quantitative evaluations discussed earlier. 

\subsection{RR Estimation}
\begin{figure}[tb]%
  \centering
  \includegraphics[width=0.48\textwidth]{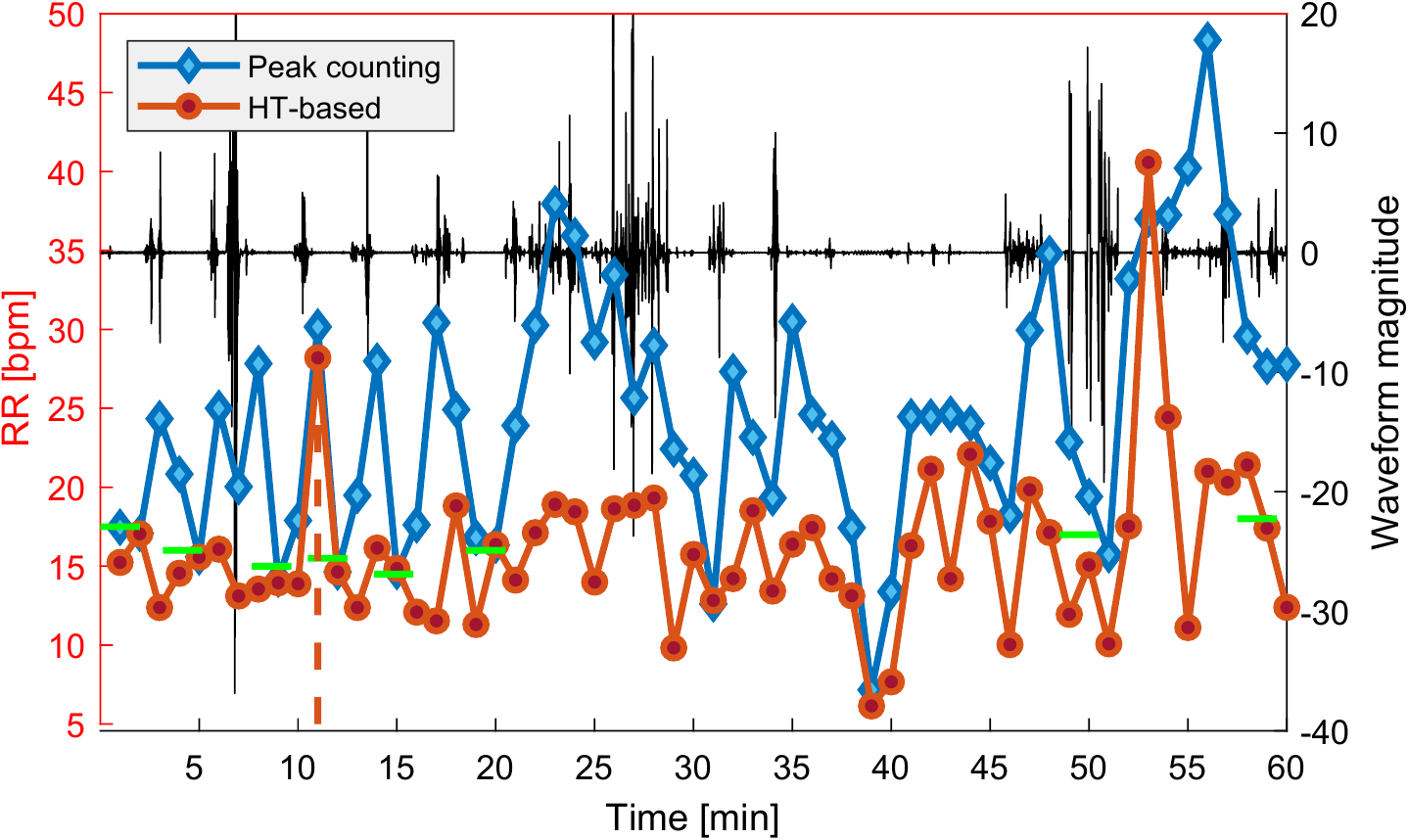}
  \caption[]{Applying both peak counting and HT to the raw waveform (black curve with the left y-axis), to extract RR for each minute (diamond and circle curves with the right y-axis). Based on the expert observations during the recording process, the participant was overall at rest in the first 20 minutes with occasional movements, and the average RR was approximately 16 bpm. Afterwards, the participant was engaged in more more movements and also took deep, slow breaths at around the 38th minute, and quick, shallow breaths at the 53rd minute, both lasting two minutes. Degraded by outliers, direct RR estimation suffers from large minute-by-minute variations, especially for the peak counting method that is more likely to over-estimate. The HT method is prone to low-energy zero-crossing noise, as illustrated by the highlighted spike at around the 11th minute. The short cyan bars are the ground-truth human observations.   
  }
  \label{fig:brraw}
\end{figure}

If RR estimation is directly applied to the raw waveform, via either peak counting or HT, the accuracy might be degraded by a higher level of noise. Fig.~\ref{fig:brraw} illustrates the RR directly extracted from a sequence lasting for an hour, updated every minute. Due to the large proportion of unreliable data caused by body movements, both peak counting and HT-based methods suffer from high errors, as reflected by the huge jumps between two neighbouring minutes. We enforced a constraint that a peak is considered as valid only if it drops on both sides by at least the standard deviation of each one-minute-long segment. However,  we still found that peak counting was more prone to outliers and showed a trend of over-estimation. HT was relatively more robust than peak counting. However, there was also a higher level of detection errors by using the HT method. Mainly, the HT method was likely to be affected by zero-crossing noise that had low energy, as shown by the spike at around the 11th minute (highlighted by the dashed line) in Fig.~\ref{fig:brraw}. 

\begin{figure}[tb]%
  \centering
  \includegraphics[width=0.48\textwidth]{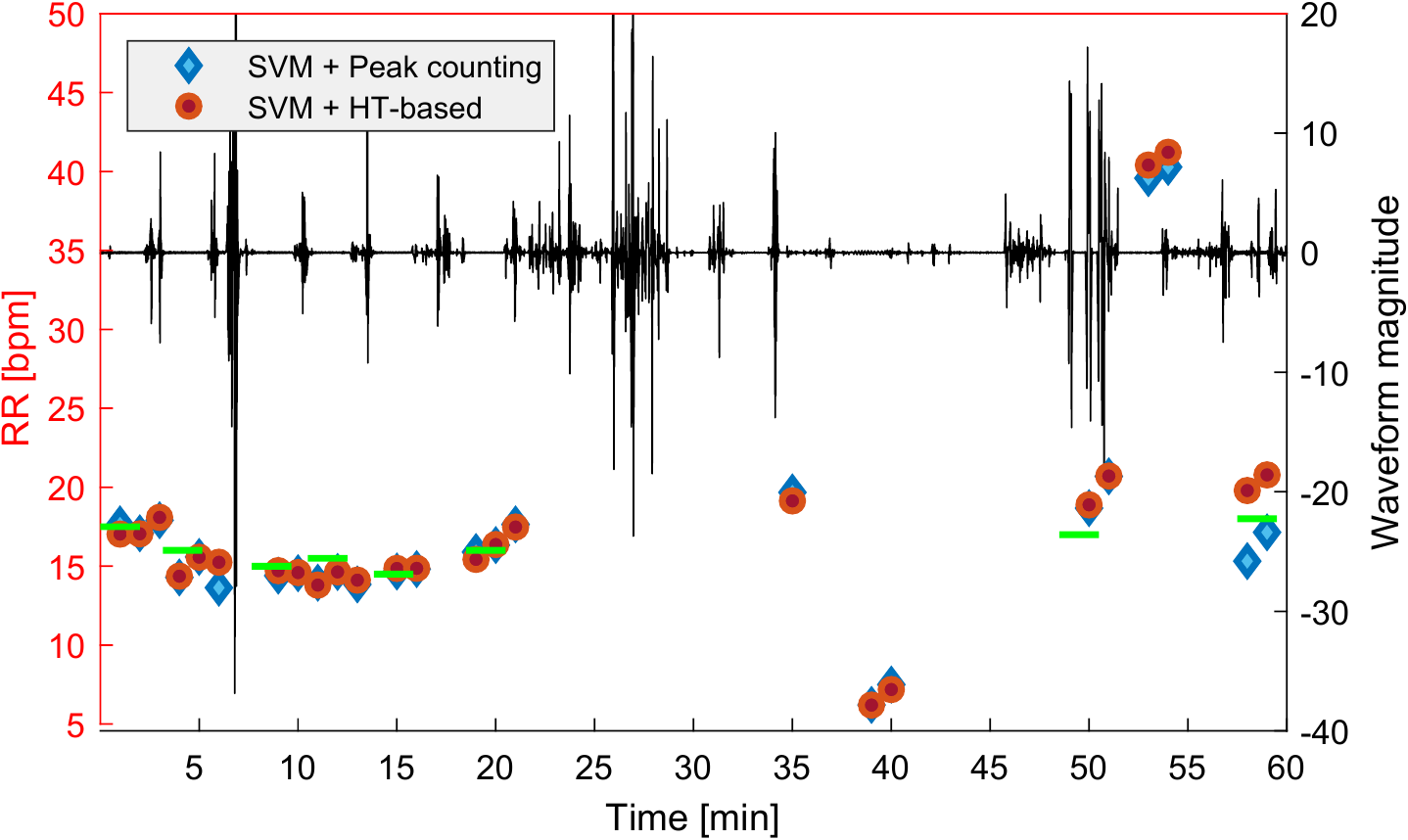}
  \caption[]{Combining the SVM model with the RR estimation for more robust respiratory rate extraction.  
  }
  \label{fig:brsvm}
\end{figure}

To resolve this problem, we combined the data reliability models built in Section~\ref{sec:exp:models} with the RR estimation methods. To start with, the trained ensembles of SVM models $\Psi_k^{\svm}(\cdot), k = 1,...,10$ were integrated. The reliability label was first predicted for each frame (second). Segments containing consecutive reliable frames were then extracted, and those that last only one or two seconds were removed. This removal of the frames is because of \textit{i)} it is very likely that no peak will be detected during those short snippets, and \textit{ii)} the slope after HT varies a lot for very short segments. If the total length of the refined segments was longer than 9~s ($15\%$ of the current minute), or if the longest valid segment lasted more than 6~s ($10\%$), we performed RR estimation by applying Eq.~(\ref{eq:peak}) and Eq.~(\ref{eq:ht}). Results on the same data as Fig.~\ref{fig:brraw} are shown in Fig.~\ref{fig:brsvm}. Due to the large proportion of movements (more than half), not every minute has an associated RR score. It is observed that during the intervals when the participant was mostly at rest, i.e. normal breathing within the first 20 minutes, deep breaths from the 38th minute, shallow breaths from the 53rd, the RR was extracted with higher accuracy, both for peak counting and HT.

We then replaced the SVM model with the first neural network model $\Psi^{\mlp}(\cdot)$. As shown in Fig.~\ref{fig:classifiers}, the MLP model had a very similar accept-reject pattern as the SVM, which was also confirmed by their similar quantitative evaluation results in error, precision and recall. As a result, MLP integration for the RR estimation yielded very similar results, as shown in Fig.~\ref{fig:brsvm}. Of all the detected RR associated with the same timestamps, SVM and MLP had an absolute average difference of 0.56 bpm and 0.42 bpm for peak counting and HT respectively. Due to their similar results, RR estimation combined with the MLP model is not plotted. 

\begin{figure}[tb]%
  \centering
  \includegraphics[width=0.48\textwidth]{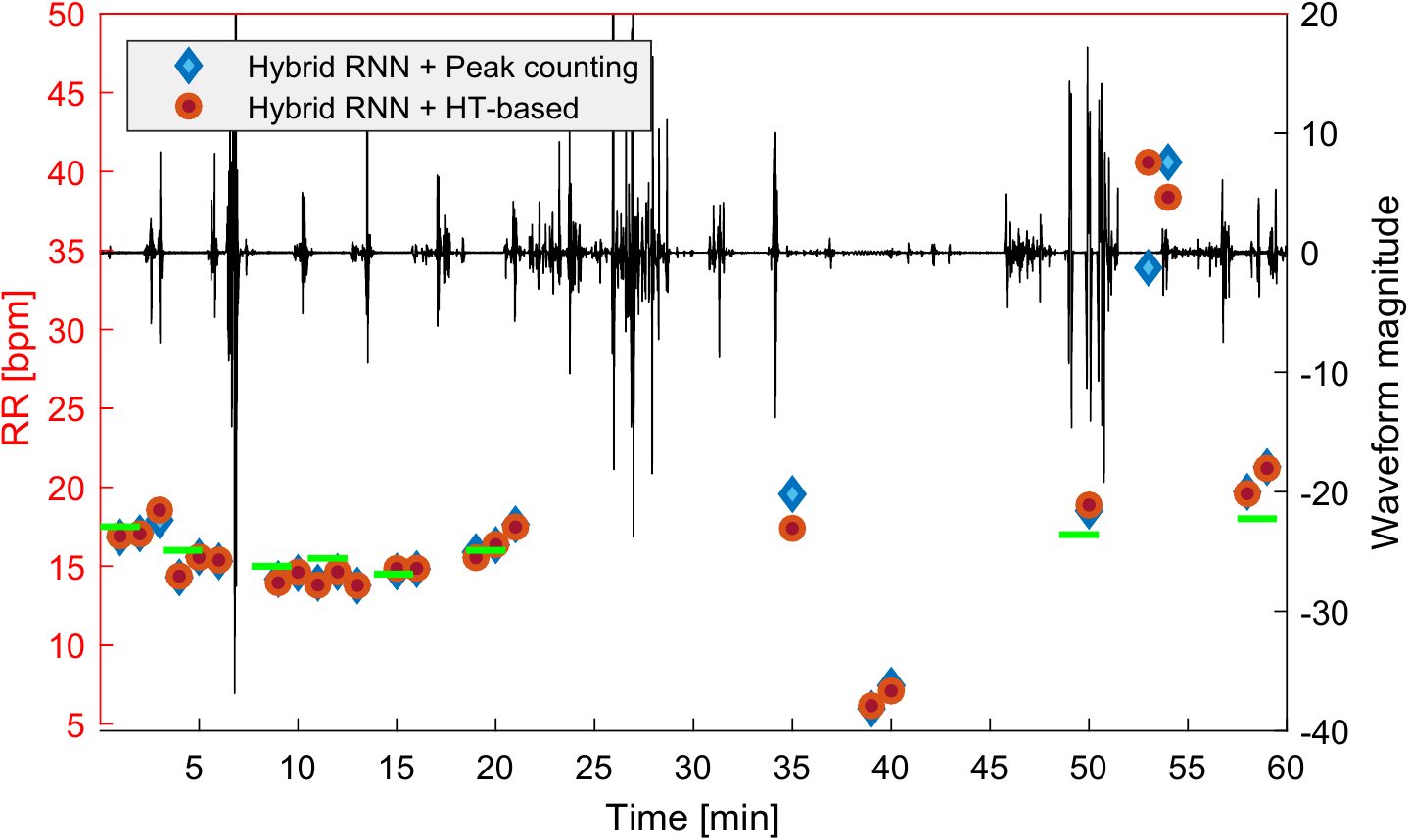}
  \caption[]{Combining the hybrid RNN model with the RR estimation. 
  }
  \label{fig:brrnn}
\end{figure}

The Hybrid RNN, on the other hand, showed a higher detection accuracy than both SVM and MLP models, as demonstrated in Fig.~\ref{fig:classifiers}. We then replaced the SVM model with the RNN model, and the final extracted RR is shown in Fig.~\ref{fig:brrnn}. As compared to Fig.~\ref{fig:brsvm}, fewer segments were associated with a RR prediction, since more frames were accurately detected as unreliable. When the participant was breathing normally at rest, i.e. the first 20 minutes, the hybrid RNN model showed almost identical RR scores to the SVM model. Interestingly, when the participant was breathing very fast at around the 53rd minute (RR=41 bpm), the peak counting only detected 34 peaks. This detection rate is due to the fact that we have enforced a constraint that only if a signal drops on both sides by at least the standard deviation of this one-minute-long segment, we consider it as a peak. This constraint is appropriate if a person is breathing normally in which the full cycles of breath have similar peak values. However, there would be several relatively more shallow breaths during the rapid breathing period, whose magnitude is very small. Thus these associated peaks were neglected. Peak counting did not introduce a significant error when the SVM model was used since frames of these shallower breaths were detected as unreliable (false-negative). In contrast to the peak counting method, HT showed more robustness to the dynamic levels of breathing. 

To quantitatively evaluate the RR estimation accuracy, we compared the estimated RR with the ground-truth RR counted by a trained observer. To mitigate the human observation error, the observer counts breaths during the time length of two minutes instead of one as used in traditional hospital measurements. In total, 28 counts lasting 56 minutes were obtained. The following evaluation metric was employed:
\begin{equation}
 e = \frac{|\text{RR}-\text{RR}^{\groundtruth}|}{|\text{RR}^{\groundtruth}|},
\end{equation}
where $\text{RR}^{\groundtruth}$ is the ground-truth RR and $|\cdot|$ is the modulus operator. The average results were reported in Table~\ref{tab:eva2}. 

\begin{table}[!ht]
\caption{Average RR estimation error $e$ from 56 minutes} 
\centering
\resizebox{0.9\columnwidth}{!}
{
  \begin{tabular}{lcccc} \hline\hline
  \rowcolor{lightgray} $ $ & Direct & SVM & MLP & Hybrid RNN \\
  \hline
  \multicolumn{4}{l}{~} \\
  Peak counting     	& 18.26\%	&6.54\%		& 5.10\% 		& 5.53\% 	\\ 
  HT-based 		& 10.43\% 	&5.48\%		& 4.40\% 		& 3.92\%  	\\
  \hline\hline
  \end{tabular}
}
\label{tab:eva2}
\end{table}

When the peak counting or HT-based RR estimation methods were directly applied to the collected waveform, i.e. ``Direct'' in Table~\ref{tab:eva2}, a high error was obtained, especially for peak counting (almost $20\%$). Even though the HT-based method showed some improvement over peak counting, there was still an error of around $10\%$. These high errors were consistent with large frame-by-frame variations in Fig.~\ref{fig:brraw}. With the proposed data reliability detection added to the RR estimation scheme, the error was significantly reduced, particularly when the Hybrid RNN classification model was combined with HT-based RR estimation. We would like to stress that despite the ground-truth is used as the gold standard. The likelihood of human observation error is, however, not taken into account.  For instance, only full cycles of breaths were counted, while partial breaths were neglected, introducing under-calculation of real RR. All the three tested models showed a significant improvement in the RR estimation accuracy, proving the effectiveness of the proposed strategy. 

To justify the practicability of our proposed RR estimation method, we did the computational complexity analysis when processing 1-minute-long data (60 frames, 600 raw waveform data points), listed as follows. The number of operations in feature extraction from the raw data is approximately $60 \times Q(2L+1)10$. The 10-fold SVM with the RBF kernel has around $10\times 60 \times n_{sv}Q$ operations, where $n_{sv}$ is the number of support vectors for each fold, which is 7876 on average in the trained SVM model when $L=3$ and $Q=15$. The computational-intensive operations in the proposed MLP mainly lie between the three fully-connected hidden layers each that each spans the size of 100, whose number is around $60\times 2 \times 100^2$. Most RNN operations exist in the two stacked and one time-distributed LSTM layers, whose number is around $60\times 80000$. Peak counting has a very low computational complexity of ${\cal O}(600)$. HT has the complexity similar to Fast Fourier Transform (FFT) plus its inverse process, which is $2{\cal O}(600\mbox{log}_{2}600)$. The total number of operations in line fitting in the HT-based method is around $6000$ when all the frames are detected as reliable. As a result, the pre-processing, peak counting-based RR estimation and HT-based RR estimation, all have very low computational complexity, which is negligible as compared to the three data reliability detection models. Particularly, SVM has the highest computational load, approximately one order higher than the hybrid RNN model, while the RNN model is around 3 times more computationally expensive than the MLP model.

\subsection{User Interface (UI)}

To inform the best clinical practices, an user-friendly web-based UI is also provided. The front-end component was implemented in JavaScript, which supports different terminals such as laptops and mobiles. Authorised users, e.g. doctors, clinicians and nurses, can log into the interactive UI to view and manage devices and patient records.

\begin{figure*}[tb]%
  \centering
  \includegraphics[width=16cm]{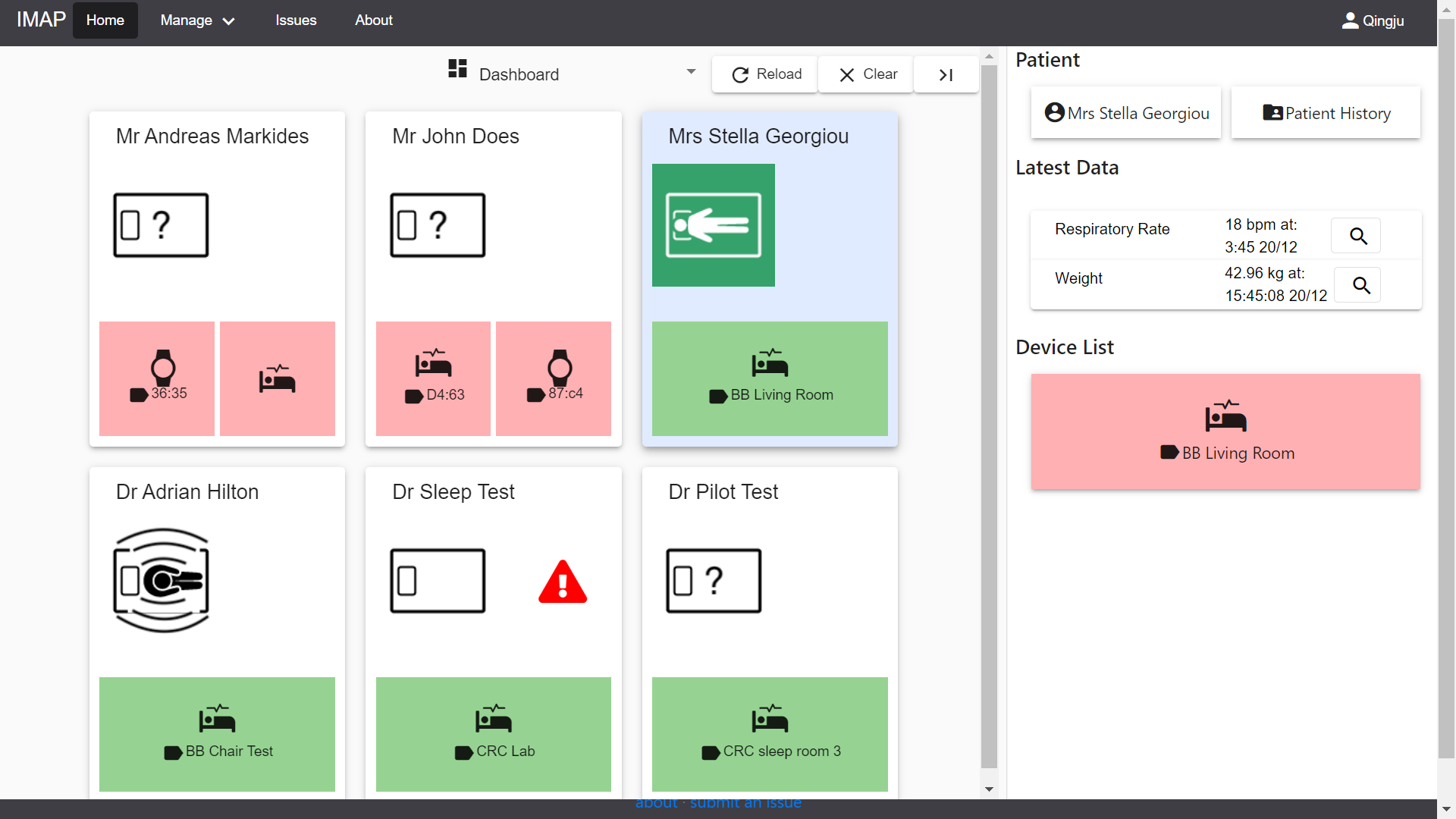}
  \caption[]{An overview UI screenshot. The thumbnail profiles on the left hand side illustrate the connection and occupation status of associated device \& patient pairs. Left-clicking one thumbnail profile will highlight that patient, and list the real-time RR and weight on the right hand side. Further information can be tracked when in the patient record.
  }
  \label{fig:ui}
\end{figure*}

Fig.~\ref{fig:ui} shows a screenshot of the UI homepage, from which it can be observed six devices were registered and assigned with patients in the database, of which two were offline (pink) and four online (green). These devices were distributed at different locations. Two beds were occupied with patients when the screenshot was captured, with one person sitting in the centre and being engaged in movements, while the other lying still in bed centre. As stressed earlier, our proposed system boosts extensibility due to its IoT structure, where other types of devices can also be integrated into our system. Fig.~\ref{fig:ui} also illustrated when wearables (the watches) that collect heart rate data are included in our system, as an extra resource of physiological information. 

\section{Conclusions}
\label{sec:conclusion}
This paper discusses an IoT-based bed sensor system for real-time health monitoring. We discuss how we have developed a method to detect the respiratory rate from measurements collected by unobtrusive sensory devices. We describe the scalability and extensibility of our approach. Bed sensor devices that used to collect physiological signals are non-contact, bringing a low awareness of operation for participants, particularly for long-term condition monitoring. 

We have developed machine learning models that enable simultaneous and robust RR estimation. The proposed models are combined with a pre-processing method to estimate the reliability and to choose the frames from raw respiratory data to contribute to the robust RR estimation. We have provided several evaluations to show the impact of the pre-processing methods and have tested and evaluated the respiratory rate against gold-standard human observations. The results show that our proposed solution can use the data collected from unobtrusive devices and provide an accurate estimation of breathing rate for a person lying on the bed, especially in the cases that the person does not move significantly on the bed.  The work has a significant impact on providing continuous and accurate RR readings in remote healthcare monitoring and in-ward hospital monitoring. The current practice is mainly based on manual observations which are resource-intensive and is also prone to human error in observations and recording. In this work, we primarily focused on the technical problems and developed an end-to-end solution to transform the raw data into reliable RR information in a continuous manner. To maximise the potential healthcare benefits and to inform the best clinical practices, the future work will focus on conducting larger scale clinical trials and real-world use-case scenarios in healthcare and hospital environments. 

We also plan to link the extracted features, i.e. RR, with more clinical markers and descriptors such as early symptom/condition detection. The future work will also involve developing machine learning methods to identify and process the extracted patterns from the data and to build predictive models. Other features collected by the bed sensor, e.g. centroid and weight, as well as features from other monitoring devices, can also be combined with RR, for more enhanced model construction and validation. In addition, the ``unreliable'' data frames that fail to contribute to RR estimation and have not been included in our methods, contain rich information that can be used to detect conditions such as periodic limb movement disorder. We will also consider using this information to extend the clinical goals of our study to other conditions.

\section*{Acknowledgements}
The authors would like to thank Dr Nikolaos Papachristou for the help with the acquisition of the dataset. This work is supported by the MinebeaMitsumi-Intec Inpatient Monitoring Activity project (IMAP). This study has received ethical approval from the Health Research Authority and Health and Care Research Wales (HCRW).

\bibliographystyle{IEEEtran}
\bibliography{minebea}

\begin{IEEEbiography}[{\includegraphics[width=1in,height=1.25in,clip,keepaspectratio]{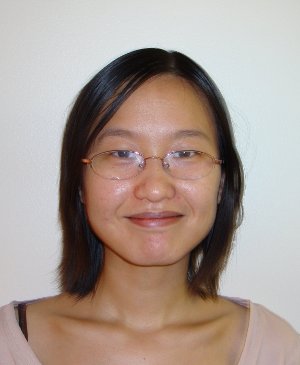}}]{Qingju Liu} received the B.Sc. degree in Electronic Information Engineering from Shandong University, Jinan, China in 2008, and the Ph.D. degree in signal processing in 2013 from the Centre for Vision, Speech and Signal Processing (CVSSP) in University of Surrey, Guildford, U.K. From October 2013 to October 2020, she worked as a research fellow in CVSSP. Since October 2020, she has been working at Huawei Cambridge Research Centre as a speech research engineer. Her current research interests include time-series analysis, IoT, audio-visual signal processing, speech recognition, neural networks and machine learning. 
\end{IEEEbiography}

\begin{IEEEbiography}[{\includegraphics[clip,width=1in,height=1.25in,keepaspectratio]{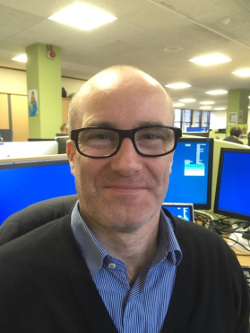}}]{Mark Kenny}, MSc, has worked for Surrey and Borders Partnership NHS Foundation Trust for 23 years, with 17 years experience as a clinician. During this period Mark has also advised and coordinated Trust services on their use of our electronic patient record systems. Over the last 5 years Mark has transitioned into a corporate role, providing a bridge between clinical and technical teams, developing healthtec projects within Digital and Innovation and Business Development Depts. This has involved building strong relationships between healthcare, academic and private industry partners, devising new treatment pathways such as within the much publicised TIHM for dementia project, and other smart ward developments. Mark is also a Research Fellow at the University of Surrey, within CVSSP. 
\end{IEEEbiography}

\begin{IEEEbiography}[{\includegraphics[trim={0 3cm 0 0},clip,width=1in,height=1.25in,keepaspectratio]{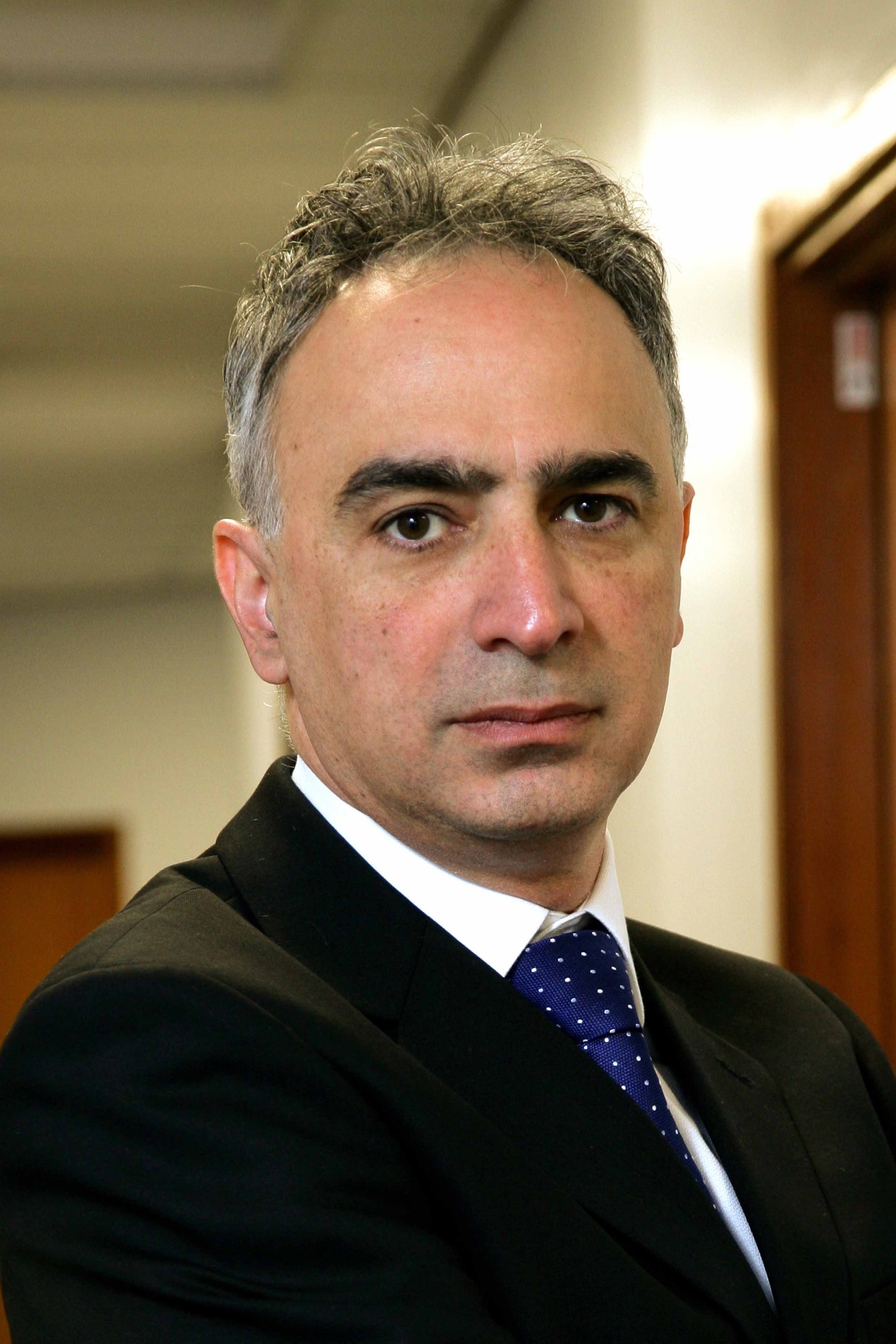}}]{Ramin Nilforooshan} is Consultant Psychiatrist at SaBP and Visiting Professor at the University of Surrey (23 publications since 2011). In addition, he is the Associate Medical Director for Research \& Development; a role which involves safety and accuracy of clinical trials and is also the Dementia Speciality Lead at Kent, Surrey and Sussex. His main research is on using IoT and AI technology for dementia care. He is the recipient of numerous HSJ awards and $>$£6.7M funding from NHS England, Innovate UK, the EU and other funders. He is the PI and CI for a number of national and international clinical trials and has considerable experience running clinical trials.
\end{IEEEbiography}

\vskip 0pt plus -1fil
\begin{IEEEbiography}[{\includegraphics[width=1in,height=1.25in,clip,keepaspectratio]{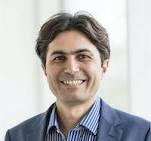}}]{Payam Barnaghi} is Professor of Machine Intelligence Applied to Medicine in the Department of Brain Sciences at Imperial College London. He is Deputy Director of Care Research and Technology Centre at the UK Dementia Research Institute (UK DRI). His research interests include machine learning, Internet of Things, semantic computing and adaptive algorithms and their applications in healthcare. He is a senior member of IEEE and a Fellow of the Higher Education Academy.
\end{IEEEbiography}

\end{document}